\documentclass[aps,showpacs]{revtex4}%
\usepackage{amsfonts}
\usepackage{latexsym}
\usepackage{epsfig}
\usepackage{amssymb}
\usepackage{amsmath}
\usepackage{graphicx}%
\begin{document}
\title{Self-sustained wormholes in modified dispersion relations}
\author{Remo Garattini}
\email{Remo.Garattini@unibg.it}
\affiliation{Universit\`{a} degli Studi di Bergamo, Facolt\`{a} di Ingegneria,}
\affiliation{Viale Marconi 5, 24044 Dalmine (Bergamo) Italy}
\affiliation{and I.N.F.N. - sezione di Milano, Milan, Italy.}
\author{Francisco S. N. Lobo}
\email{flobo@cii.fc.ul.pt}
\affiliation{Centro de Astronomia e Astrof\'{\i}sica da Universidade de Lisboa,}
\affiliation{Campo Grande, Ed. C8 1749-016 Lisboa, Portugal}
\date{\today }

\begin{abstract}
In this work, we consider the possibility that wormhole geometries are
sustained by their own quantum fluctuations, in the context of modified
dispersion relations. More specifically, the energy density of the graviton
one-loop contribution to a classical energy in a wormhole background is
considered as a self-consistent source for wormholes. In this semi-classical
context, we consider specific choices for the Rainbow's functions and find
solutions for wormhole geometries in the cis- planckian and trans-planckian
regimes. In the former regime, the wormhole spacetimes are not asymptotically
flat and need to be matched to an exterior vacuum solution. In the latter
trans-planckian regime, we find that the quantum corrections are exponentially
suppressed, which provide asymptotically flat wormhole geometries with a 
constant shape function, i.e., $b(r)=r_{t}$,
where $r_{t}$ is the wormhole throat. In addition to this analysis, we also
fix the geometry by considering the behaviour of a specific shape function
through a variational approach which imposes a local analysis to the problem
at the wormhole throat. We further explore the respective parameter range of the
Rainbow's functions, and find a good agreement with previous work.

\end{abstract}

\pacs{04.60.-m, 04.20.Jb}
\maketitle

\section{Introduction}
\label{p1}

Modified Dispersion Relations (MDR) are a distortion of the spacetime metric
at energies comparable to the Planck energy. A general formalism, denoted as
deformed or doubly special relativity \cite{GAC} was developed in
\cite{MagSmo} in order to (i) preserve the relativity of inertial frames, (ii)
maintain the Planck energy invariant and (iii) impose that in the limit
$E/E_{P} \rightarrow0$ the speed of a massless particle tends to a universal
constant, $c$, which is the same for all inertial observers. In particular, in
curved spacetime the formalism imposes that the relationship between the
energy and momentum of a massive particle $m$ in special relativity are
modified at the Planck scale \cite{GAC,MagSmo}, i.e., $E^{2} g_{1}^{2}%
(E/E_{P})-p^{2} g_{2}^{2}(E/E_{P})=m^{2}$, where the two unknown functions
$g_{1} (E/E_{P})$ and $g_{2}(E/E_{P})$, denoted as the \textit{Rainbow's
functions}, have the following property
\begin{equation}
\lim_{E/E_{P}\rightarrow0}g_{1}\left(  E/E_{P}\right)  =1\qquad\text{and}%
\qquad\lim_{E/E_{P}\rightarrow0}g_{2}\left(  E/E_{P}\right)  =1. \label{lim}%
\end{equation}

One interesting aspect of Rainbow's functions is the possibility of
regularizing the divergent behaviour of some field quantities such as the energy
density. This has been done in Ref. \cite{RGGM} with the following choice for
the Rainbow's functions
\begin{equation}
g_{1}\left(  E/E_{P}\right)  =\sum_{i=0}^{n}\beta_{i}\frac{E^{i}}{E_{P}^{i}%
}\exp\left(  -\alpha\frac{E^{2}}{E_{P}^{2}}\right)  ,\qquad g_{2}\left(
E/E_{P}\right)  =1;\qquad\alpha>0,\beta_{i}\in\mathbb{R}.
\label{Remofunctions}%
\end{equation}
More specifically, in Ref. \cite{RGGM}, motivated by the promising results
obtained in the application of Gravity's Rainbow to black hole entropy
\cite{RemoPLB}, it was shown that for an appropriate choice of the functions
$g_{1} (E/EP )$ and $g_{2} (E/EP )$, the UV divergences of the Zero Point Energy
disappear. It is interesting to note that every choice for the Rainbow's
functions is restricted by convergence criteria, namely a pure polynomial
cannot be used without the reintroduction of a regularization and a
renormalization process.

In fact, MDR have been considered in a wide variety of contexts, such as in
cosmology \cite{MDRcosmo} and black hole physics \cite{MDRbh}. For instance,
the \textit{rainbow} version of the Schwarzschild line element is given by
\cite{MagSmo}%
\begin{equation}
ds^{2}=-\left(  1-\frac{2MG(0) }{r}\right)  \frac{d\tilde{t}^{2}}{g_{1}%
^{2}\left(  E/E_{P}\right)  }+\frac{d\tilde{r}^{2}}{\left(  1-\frac{2MG(0)
}{r}\right)  g_{2}^{2}\left(  E/E_{P}\right)  }+\frac{\tilde{r}^{2}}{g_{2}%
^{2}\left(  E/E_{P}\right)  }\left(  d\theta^{2}+\sin^{2}\theta d\phi
^{2}\right)  \,. \label{line}%
\end{equation}
In particular, note that the position of the horizon, in fixed and energy
independent coordinates, is at the usual fixed coordinate $2MG(0)$. However,
it was shown that the area of the horizon is energy dependent \cite{MagSmo}.

In this paper, we use the scheme given by Rainbow's function$\left(
\ref{Remofunctions}\right)  $, in order to verify if the MDR provide
interesting solutions for wormhole geometries. The latter are hypothetical
tunnels in spacetime \cite{Morris,Visser} and have been studied in a plethora of
environments such as dark energy models \cite{phantomWH}, modified theories of
gravity \cite{modgrav}, observational signatures using thin accretion disks
\cite{Harko:2008vy}, and in the semi-classical regime
\cite{semiclass,Garattini,semiclassWH}, amongst many other contexts. In the
semi-classical regime, the possibility that these wormholes are sustained by
their own quantum fluctuations was considered in \cite{Garattini,semiclassWH}.
The graviton one loop contribution to a classical energy in a wormhole
background was taken into account, and a variational approach with Gaussian
trial wave functionals was used. A zeta function regularization was involved
to handle the divergences and a renormalization procedure was introduced. Thus,
the finite one loop energy was considered as a \textit{self-consistent} source
for traversable wormholes. The aim of this work is to consider the
possibility that wormhole geometries be sustained by their own quantum
fluctuations, in the context of modified dispersion relations.

This paper is organized in the following manner: In Section \ref{p2} we
briefly outline the formalism of the classical term in Rainbow's Gravity and
the one loop energy in a spherically symmetric background. In Section
\ref{p3}, we consider specific Rainbow's functions and find solutions of
self-sustained wormhole geometries in the context of MDR. In Section \ref{p4},
we fix the geometry by considering the behaviour of a specific metric function
$b(r)$, through a variational approach, at the wormhole throat, $r_{t}$. In
Section \ref{p5}, we conclude.

\section{The classical term in Rainbow's Gravity and the one loop energy in a
spherically symmetric background}
\label{p2}

Consider a wormhole spacetime described by the following line element
\begin{equation}
ds^{2}=-N^{2}\left(  r\right)  \frac{dt^{2}}{g_{1}^{2}\left(  E/E_{P}\right)
}+\frac{dr^{2}}{\left(  1-\frac{b\left(  r\right)  }{r}\right)  g_{2}%
^{2}\left(  E/E_{P}\right)  }+\frac{r^{2}}{g_{2}^{2}\left(  E/E_{P}\right)
}\left(  d\theta^{2}+\sin^{2}\theta d\phi^{2}\right)  \,, \label{dS}%
\end{equation}
where $N(r)$ is the lapse function and $b(r)$ is denoted the shape function,
as can be shown by embedding diagrams, it determines the shape of the wormhole
\cite{Morris}. For simplicity, we consider $N(r)=1$ throughout this work. A
fundamental property of traversable wormholes is the flaring out condition of
the throat, given by $(b-b^{\prime}r)/b^{2}>0$ \cite{Morris}. Note that the
condition $1-b/r>0$ is also imposed. To be a wormhole solution the following
conditions need to be satisfied at the throat: $b(r_{t})=r_{t}$ and $b^{\prime
}(r_{t})<1$; the latter follows from the flaring out condition. Asymptotic
flatness imposes $b(r)/r\rightarrow0$ as $r\rightarrow+\infty$. However, one
may also construct solutions by matching the interior solution to an exterior
vacuum spacetime, at a junction interface, much in the spirit of
\cite{matching}.

In the analysis outlined below, we consider the graviton one loop contribution
to a classical energy in a wormhole background \cite{Garattini}, where the
classical energy is given by
\begin{equation}
H_{\Sigma}^{(0)}=\int_{\Sigma}\,d^{3}x\,\mathcal{H}^{(0)}=-\frac{1}{16\pi
G}\int_{\Sigma}\,d^{3}x\,\sqrt{g}\,R\,=-\frac{1}{2G}\int_{r_{0}}^{\infty
}\,\frac{dr\,r^{2}}{\sqrt{1-b(r)/r}}\,\frac{b^{\prime}(r)}{r^{2}g_{2}\left(
E/E_{P}\right)  }\,, \label{classical}%
\end{equation}
and the background field super-hamiltonian, $\mathcal{H}^{(0)}$, is integrated
on a constant time hypersurface. Note that the graviton one loop contribution
to a classical energy contribution is evaluated through a variational approach
with Gaussian trial wave functionals, and the divergences are treated with a
zeta function regularization. Using a renormalization procedure, the finite
one loop energy was considered a self-consistent source for a traversable
wormhole (we refer the reader to \cite{Garattini} for details).

In the following, we consider $g_{ij}=\bar{g}_{ij}+h_{ij}$, where $h_{ij}$ is
the quantum fluctuation around the background metric $\bar{g}_{ij}$. We shall
also take into account the total regularized one loop energy given by
\begin{equation}
E^{TT}=-\frac{1}{2}\sum_{\tau}\frac{g_{1}\left(  E/E_{P}\right)  }{g_{2}%
^{2}\left(  E/E_{P}\right)  }\left[  \sqrt{E_{1}^{2}\left(  \tau\right)
}+\sqrt{E_{2}^{2}\left(  \tau\right)  }\right]  \,, \label{ETT}%
\end{equation}
where $E_{i}^{2}\left(  \tau\right)  >0$, and $E_{i}$ are the eigenvalues of
\begin{equation}
\left(  \hat{\bigtriangleup}_{L\!}^{m}\!{}\;{h}^{\bot}\right)  _{ij}\!{}%
=\frac{E^{2}}{g_{2}^{2}\left(  E/E_{P}\right)  }{h}_{ij}^{\bot}\,. \label{EE}%
\end{equation}
${h}^{\bot}$ is the traceless-transverse component of the perturbation
\cite{Vassilevich,Quad} and
\begin{equation}
\left(  \hat{\bigtriangleup}_{L\!}^{m}\!{}\;h^{\bot}\right)  _{ij}=\left(
\bigtriangleup_{L\!}\!{}\;h^{\bot}\right)  _{ij}-4R{}_{i}^{k}\!{}%
\;h_{kj}^{\bot}+\text{ }^{3}R{}\!{}\;h_{ij}^{\bot}\,, \label{M Lichn}%
\end{equation}
where $\bigtriangleup_{L}$ is the Lichnerowicz operator defined by%
\begin{equation}
\left(  \bigtriangleup_{L}\;h\right)  _{ij}=\bigtriangleup h_{ij}%
-2R_{ikjl}\,h^{kl}+R_{ik}\,h_{j}^{k}+R_{jk}\,h_{i}^{k}, \label{DeltaL}%
\end{equation}
with $\bigtriangleup=-\nabla^{a}\nabla_{a}$. Rather than present all 
the intricate details here, we refer the reader to Ref.
\cite{RGGM} for a detailed discussion on these issues.

Using the Regge and Wheeler representation \cite{Regge Wheeler}, the
eigenvalue equation $\left(  \ref{EE}\right)  $ can be reduced to
\begin{equation}
\left[  -\frac{d^{2}}{dx^{2}}+\frac{l\left(  l+1\right)  }{r^{2}}+m_{i}%
^{2}\left(  r\right)  \right]  f_{i}\left(  x\right)  =\frac{E_{i,l}^{2}%
}{g_{2}^{2}\left(  E/E_{P}\right)  }f_{i}\left(  x\right)  \qquad(i=1,2)\,,
\label{p34}%
\end{equation}
where we have used reduced fields of the form $f_{i}\left(  x\right)
=F_{i}\left(  x\right)  /r$, and have defined, for simplicity, two
$r-$dependent effective masses $m_{1}^{2}\left(  r\right)  $ and $m_{2}%
^{2}\left(  r\right)  $ given by
\begin{equation}
\left\{
\begin{array}
[c]{c}%
m_{1}^{2}\left(  r\right)  =\frac{6}{r^{2}}\left(  \,1-\frac{b(r)}{r}\right)
-\frac{3b^{\prime}(r)}{2r^{2}}+\frac{3b(r)}{2r^{3}}\,\\
\\
m_{2}^{2}\left(  r\right)  =\frac{6}{r^{2}}\left(  \,1-\frac{b(r)}{r}\right)
-\frac{b^{\prime}(r)}{2r^{2}}-\frac{3b(r)}{2r^{3}}%
\end{array}
\right.  \quad\left(  r\equiv r\left(  x\right)  \right)  . \label{masses}%
\end{equation}

Taking into account the WKB approximation, from Eq. $\left(  \ref{p34}\right)
$ we extract two $r-$dependent radial wave numbers given by
\begin{equation}
k_{i}^{2}\left(  r,l,\omega_{i,nl}\right)  =\frac{E_{i,nl}^{2}}{g_{2}%
^{2}\left(  E/E_{P}\right)  }-\frac{l\left(  l+1\right)  }{r^{2}}-m_{i}%
^{2}\left(  r\right)  \qquad(i=1,2)\,. \label{kTT}%
\end{equation}

It is useful to use the WKB method implemented by `t Hooft in the brick wall
problem \cite{tHooft}, by counting the number of modes with frequency less
than $\omega_{i}$, $i=1,2$. This is given by%
\begin{equation}
\tilde{g}\left(  E_{i}\right)  =\int_{0}^{l_{\max}}\nu_{i}\left(
l,E_{i}\right)  \left(  2l+1\right)  dl\,, \label{p41}%
\end{equation}
where $\nu_{i}\left(  l,E_{i}\right)  $, $i=1,2$, is the number of nodes in
the mode with $\left(  l,E_{i}\right)  $, such that $\left(  r\equiv r\left(
x\right)  \right)  $
\begin{equation}
\nu_{i}\left(  l,E_{i}\right)  =\frac{1}{\pi}\int_{-\infty}^{+\infty}%
dx\sqrt{k_{i}^{2}\left(  r,l,E_{i}\right)  }\,. \label{p42}%
\end{equation}
Note that the integration with respect to $x$ and $l_{\max}$ is taken over
those values which satisfy $k_{i}^{2}\left(  r,l,E_{i}\right)  \geq0,$
$i=1,2$. With the help of Eqs. $\left(  \ref{p41}\right)  $ and $\left(
\ref{p42}\right)  $, the self-sustained traversable wormhole equation becomes%
\begin{equation}
H_{\Sigma}^{(0)}=-\frac{1}{\pi}\sum_{i=1}^{2}\int_{0}^{+\infty}E_{i}%
\frac{g_{1}\left(  E/E_{P}\right)  }{g_{2}^{2}\left(  E/E_{P}\right)  }%
\frac{d\tilde{g}\left(  E_{i}\right)  }{dE_{i}}dE_{i}\,. \label{tot1loop}%
\end{equation}
The explicit evaluation of the density of states yields%
\begin{align}
\frac{d\tilde{g}(E_{i})}{dE_{i}}  &  =\int\frac{\partial\nu(l{,}E_{i}%
)}{\partial E_{i}}(2l+1)dl\nonumber\\
&  =\frac{1}{\pi}\int_{-\infty}^{+\infty}dx\int_{0}^{l_{\max}}\frac
{(2l+1)}{\sqrt{k^{2}(r,l,E)}}\frac{d}{dE_{i}}\left(  \frac{E_{i}^{2}}%
{g_{2}^{2}\left(  E/E_{P}\right)  }-m_{i}^{2}\left(  r\right)  \right)
dl\nonumber\\
&  =\frac{2}{\pi}\int_{-\infty}^{+\infty}dxr^{2}\frac{d}{dE_{i}}\left(
\frac{E_{i}^{2}}{g_{2}^{2}\left(  E/E_{P}\right)  }-m_{i}^{2}\left(  r\right)
\right)  \sqrt{\frac{E_{i}^{2}}{g_{2}^{2}\left(  E/E_{P}\right)  }-m_{i}%
^{2}\left(  r\right)  }\nonumber\\
&  =\frac{4}{3\pi}\int_{-\infty}^{+\infty}dxr^{2}\frac{d}{dE_{i}}\left(
\frac{E_{i}^{2}}{g_{2}^{2}\left(  E/E_{P}\right)  }-m_{i}^{2}\left(  r\right)
\right)  ^{\frac{3}{2}}. \label{states}%
\end{align}

Finally, plugging expression $\left(  \ref{states}\right)  $ into Eq. $\left(
\ref{tot1loop}\right)  $ and taking into account the energy density, we obtain
the following self-sustained equation
\begin{equation}
\frac{1}{2G}\,\frac{b^{\prime}(r)}{r^{2}g_{2}\left(  E/E_{P}\right)  }%
=\frac{2}{3\pi^{2}}\left(  I_{1}+I_{2}\right)  \,, \label{LoverG}%
\end{equation}
which will play an important role in the analysis below. The integrals $I_{1}$
and $I_{2}$ are defined as
\begin{equation}
I_{1}=\int_{E^{\ast}}^{\infty}E\frac{g_{1}\left(  E/E_{P}\right)  }{g_{2}%
^{2}\left(  E/E_{P}\right)  }\frac{d}{dE}\left(  \frac{E^{2}}{g_{2}^{2}\left(
E/E_{P}\right)  }-m_{1}^{2}\left(  r\right)  \right)  ^{\frac{3}{2}}dE\,,
\label{I1}%
\end{equation}
and%
\begin{equation}
I_{2}=\int_{E^{\ast}}^{\infty}E\frac{g_{1}\left(  E/E_{P}\right)  }{g_{2}%
^{2}\left(  E/E_{P}\right)  }\frac{d}{dE}\left(  \frac{E^{2}}{g_{2}^{2}\left(
E/E_{P}\right)  }-m_{2}^{2}\left(  r\right)  \right)  ^{\frac{3}{2}}dE\,,
\label{I2}%
\end{equation}
respectively. Note that $E^{\ast}$ is the value which annihilates the argument of the root.

In $I_{1}$ and $I_{2}$ we have included an additional $4\pi$ factor coming
from the angular integration and we have assumed that the effective mass does
not depend on the energy $E$. To further proceed, we can see what happens to
expression $\left(  \ref{LoverG}\right)  $ for some specific forms of
$g_{1}\left(  E/E_{P}\right)  $ and $g_{2}\left(  E/E_{P}\right)  $. It is
immediate to see that integrals $I_{1}$ and $I_{2}$ can be solved when
$g_{2}(E/E_{P})=g_{1}(E/E_{P})$. However the classical term keeps a dependence
on the function $g_{2}(E/E_{P})$ that cannot be eliminated except for the
simple case of $g_{2}(E/E_{P})=1$. Therefore we will consider different models
regulated by the Rainbow's function $g_{1}\left(  E/E_{P}\right)  $, of the
form given by Eq. (\ref{Remofunctions}), to analyse the effect on the form of
the shape function $b(r)$.

\section{Examples}
\label{p3}

\subsection{Specific case: $g_{1}\left(  E/E_{P}\right)  =\exp(-\alpha
\frac{E^{2}}{E_{P}^{2}}),\qquad g_{2}\left(  E/E_{P}\right)  =1$}
\label{p3a}

Following Ref. \cite{RGGM}, we consider the following choice for the Rainbow's
functions
\begin{equation}
g_{1}\left(  E/E_{P}\right)  =\exp(-\alpha\frac{E^{2}}{E_{P}^{2}}),\qquad
g_{2}\left(  E/E_{P}\right)  =1;\qquad\alpha>0\in\mathbb{R}. \label{g1g2}%
\end{equation}
Thus, the graviton contribution terms $(\ref{I1})$ and $(\ref{I2})$ yield the
following relationships
\begin{equation}
I_{1}=3\int_{\sqrt{m_{1}^{2}\left(  r\right)  }}^{\infty}\exp(-\alpha
\frac{E^{2}}{E_{P}^{2}})E^{2}\sqrt{E^{2}-m_{1}^{2}\left(  r\right)  }dE \,,
\label{I11}%
\end{equation}
and%
\begin{equation}
I_{2}=3\int_{\sqrt{m_{2}^{2}\left(  r\right)  }}^{\infty}\exp(-\alpha
\frac{E^{2}}{E_{P}^{2}})E^{2}\sqrt{E^{2}-m_{2}^{2}\left(  r\right)  }dE\,,
\label{I22}%
\end{equation}
respectively. Using the general results outlined in Appendix \ref{appe} for
the two integrals $I_{1}$ and $I_{2}$, Eq. $(\ref{LoverG})$ can be rearranged
in the following way%
\begin{equation}
\frac{1}{2G}\,\frac{b^{\prime}(r)}{r^{2}}=\frac{E_{P}^{4}}{2\pi^{2}}\left[
\frac{x_{1}^{2}}{\alpha}\exp\left(  -\frac{\alpha x_{1}^{2}}{2}\right)
K_{1}\left(  \frac{\alpha x_{1}^{2}}{2}\right)  +\frac{x_{2}^{2}}{\alpha}%
\exp\left(  \frac{\alpha x_{2}^{2}}{2}\right)  K_{1}\left(  \frac{\alpha
x_{2}^{2}}{2}\right)  \right]  , \label{bp}%
\end{equation}
where $x_{1}=\sqrt{m_{1}^{2}\left(  r\right)  /E_{P}^{2}}$ , $x_{2}%
=\sqrt{m_{2}^{2}\left(  r\right)  /E_{P}^{2}}$ and $K_{1}\left(  x\right)  $
is a modified Bessel function of order 1. Note that it is extremely difficult
to extract any useful information from this relationship, so that in the
following we consider two regimes, namely the cis-planckian regime, where
$x_{i}\ll1$ ($i=1,2$), and the trans-planckian r\'{e}gime, where $x_{i}\gg1$.

In the cis-planckian regime, with the approximation $x_{1}\ll1$ and $x_{2}%
\ll1$, and expanding the right hand side of Eq. $\left(  \ref{bp}\right)  $,
we find that the leading term is given by
\begin{equation}
\frac{1}{2G}\,\frac{b^{\prime}(r)}{r^{2}}\simeq\frac{E_{P}^{4}}{2\pi^{2}%
}\left[  \frac{4}{\alpha^{2}}-\frac{1}{\alpha}\left(  x_{1}^{2}+x_{2}%
^{2}\right)  +O\left(  x_{1}^{4}+x_{2}^{4}\right)  \right]  \,.
\end{equation}
Substituting the factors $x_{i}=\sqrt{m_{i}^{2}\left(  r\right)  /E_{P}^{2}}$
($i=1,2$) in the latter, provides the following relationship
\begin{equation}
\frac{1}{2G}\,\frac{b^{\prime}(r)}{r^{2}}=\frac{E_{P}^{4}}{2\pi^{2}}\left[
\frac{4}{\alpha^{2}}-\frac{12}{\alpha r^{2}E_{P}^{2}}\left(  \,1-\frac
{b(r)}{r}\right)  +\frac{2b^{\prime}(r)}{\alpha E_{P}^{2}r^{2}}\right]  ,
\end{equation}
which can be rearranged to give%
\begin{equation}
b^{\prime}(r)=\frac{1}{\pi^{2}}\left[  \frac{4r^{2}}{\alpha^{2}G}-\frac
{12}{\alpha}\left(  \,1-\frac{b(r)}{r}\right)  +\frac{2b^{\prime}(r)}{\alpha
}\right]  \,,
\end{equation}
where we have used the definition $G=E_{P}^{-2}=l_{P}^{2}$. Restricting our
attention to the dominant term only, we find that
\begin{equation}
b(r)=r_{t}+\frac{E_{P}^{2}}{3\pi^{2}\alpha^{2}}\left(  r^{3}-r_{t}^{3}\right)
\,,
\end{equation}
which does not represent an asymptotically flat wormhole geometry, as the
condition $b(r)/r\rightarrow0$ when $r\rightarrow+\infty$, is not satisfied.
However, for these cases, one may in principle match these interior wormhole
solutions with an exterior vacuum Schwarzschild spacetime. We shall not
proceed with the matching analysis in this paper, but we refer the reader to
\cite{matching} for further details.

On the other hand, in the trans-planckian regime, i.e., $x_{1}\gg1$ and
$x_{2}\gg1$, we obtain the following approximation
\begin{equation}
\frac{1}{2G}\,\frac{b^{\prime}(r)}{r^{2}}\simeq\frac{E_{P}^{4}}{8\sqrt
{\alpha^{3}\pi^{3}}}\left[  \exp\left(  -\alpha x_{1}^{2}\right)
x_{1}+O\left(  \frac{1}{x_{1}}\right)  +\exp\left(  -\alpha x_{2}^{2}\right)
x_{2}+O\left(  \frac{1}{x_{2}}\right)  \right]  \,.
\end{equation}
Note that in this regime, the asymptotic expansion is dominated by the
Gaussian exponential so that the quantum correction vanishes. Thus, the only
solution is $b^{\prime}(r)=0$ and consequently we have a constant shape
function, namely, $b(r)=r_{t}$. In summary, in the trans-planckian regime provided by the Rainbow's 
function $(\ref{g1g2})$, we verify that the self-sustained
equation $\left(  \ref{LoverG}\right)  $ permits asymptotically flat wormhole
solutions with a constant shape function given by $b(r)=r_{t}$.

\subsection{Specific case: $g_{1}\left(  E/E_{P}\right)  =\left(  1+\beta
\frac{E}{E_{P}}\right)  \exp(-\alpha\frac{E^{2}}{E_{P}^{2}}),\qquad
g_{2}\left(  E/E_{P}\right)  =1$}
\label{p3b}

Another interesting choice for the Rainbow's functions is the following
\cite{RGGM},
\begin{equation}
g_{1}\left(  E/E_{P}\right)  =\left(  1+\beta\frac{E}{E_{P}}\right)
\exp(-\alpha\frac{E^{2}}{E_{P}^{2}}),\qquad g_{2}\left(  E/E_{P}\right)
=1;\qquad\alpha>0,\beta\in\mathbb{R}.
\end{equation}
For these specific functions, we once again use the general results in
Appendix \ref{appe} so that the two integrals $I_{1}$ and $I_{2}$, given by
Eqs. $\left(  \ref{I1}\right)  $-$\left(  \ref{I2}\right)  $, take the
following form
\begin{equation}
I_{1,2}=\frac{E_{P}^{4}}{2\pi^{2}}\left[  \frac{x_{1,2}^{2}}{\alpha}%
\exp\left(  -\frac{\alpha x_{1,2}^{2}}{2}\right)  K_{1}\left(  \frac{\alpha
x_{1,2}^{2}}{2}\right)  +\beta\frac{\sqrt{\pi}}{{\alpha}^{\frac{3}{2}}}\left(
x_{1,2}^{2}+\frac{3}{2{\alpha}}\right)  \exp\left(  -\alpha x_{1,2}%
^{2}\right)  \right]  \,.
\end{equation}
where once again, we have defined for notational simplicity $x_{1,2}%
=\sqrt{m_{1,2}^{2}\left(  r\right)  /E_{P}^{2}}$. As in the previous example,
it is extremely difficult to extract any useful information from these
relationships, so that we consider the two regimes, i.e., the cis-planckian
regime, where $x_{i}\ll1$ ($i=1,2$), and the trans-planckian regime, where
$x_{i}\gg1$, respectively.

In the cis-planckian regime, where $x_{i}\ll1$, the self-sustained equation
$\left(  \ref{LoverG}\right)  $ takes the following form
\begin{equation}
\frac{1}{2G}\,\frac{b^{\prime}(r)}{r^{2}}=\frac{E_{P}^{4}}{2{\alpha}\pi^{2}%
}\left[  \frac{2}{{\alpha}}+{\frac{3\sqrt{\pi}\beta}{2{\alpha}^{3/2}}%
-\frac{2\sqrt{{\alpha}}+\sqrt{\pi}\beta}{2{\alpha}^{1/2}}}\left(  {{x}_{1}%
^{2}+{x}_{2}^{2}}\right)  +O\left(  x^{4}\right)  \right]  , \label{AsL}%
\end{equation}
which leads to
\begin{equation}
b^{\prime}(r)=\frac{E_{P}^{2}}{{\alpha}\pi^{2}}\left[  {\frac{4\sqrt{{\alpha}%
}+3\sqrt{\pi}\beta}{2{\alpha}^{3/2}}r^{2}-}\frac{2\sqrt{{\alpha}}+\sqrt{\pi
}\beta}{{\alpha}^{1/2}}\left(  \frac{6}{E_{P}^{2}}\left(  \,1-\frac{b(r)}%
{r}\right)  -\frac{b^{\prime}(r)}{E_{P}^{2}}\right)  \right]  .
\end{equation}
Note that the term proportional to $r^{2}$ leads to a non-asymptotically flat
wormhole spacetime, so that it is useful to do away with this term. To this
effect, it is straightforward to see that for $\beta=-\frac{4}{3}\sqrt
{\alpha/\pi}$, one arrives at
\begin{equation}
b^{\prime}(r)=-\frac{2}{3\alpha\pi^{2}}\left[
6\left(  \,1-\frac{b(r)}{r}\right)  -b^{\prime}(r) \right]  ,
\end{equation}
which provides the following solution
\begin{equation}
b(r)=\frac{1}{3\alpha\pi^{2}-12\pi^{2}-2}\left[  \left(  3\alpha\pi
^{2}-2\right)  r_{t}{}\left(  \frac{r}{r_{t}}\right)  ^{\frac{12\pi^{2}%
}{3\alpha\pi^{2}-2}}-12\pi^{2}r\right]  ,
\end{equation}
where we have used the condition $b\left(  r_{t}\right)  =r_{t}$. It is a
simple matter to verify that this solution is not asymptotically flat,
however, as in the previous example one may match the interior solution to an
exterior solution (once again, we refer the reader to \cite{matching} for
further details).

On the other hand, for the trans-planckian regime, $x_{i}\gg1$, the asymptotic
series becomes
\begin{equation}
\frac{1}{2G}\,\frac{b^{\prime}(r)}{r^{2}}\simeq\frac{E_{P}^{4}}{8\sqrt
{\alpha^{3}\pi^{3}}}\left[  \left(  x_{1}+\beta x_{1}^{2}\right)  \exp\left(
-\alpha x_{1}^{2}\right)  +O\left(  \frac{1}{x_{1}}\right)  +\left(
x_{2}+\beta x_{2}^{2}\right)  \exp\left(  -\alpha x_{2}^{2}\right)  +O\left(
\frac{1}{x_{2}}\right)  \right]  .
\end{equation}
Once again, we find that the quantum corrections are exponentially suppressed,
and as in the previous example leads to a constant shape function,
$b(r)=r_{t}$.

\subsection{Specific case: $g_{1}\left(  E/E_{P}\right)  =\left(  1+\beta
\frac{E}{E_{P}}+\gamma\frac{E^{2}}{E_{P}^{2}}\right)  \exp(-\alpha\frac{E^{2}%
}{E_{P}^{2}}),\qquad g_{2}\left(  E/E_{P}\right)  =1$}
\label{p3c}

Finally, consider now the following choices for the Rainbow's functions given
by
\begin{equation}
g_{1}\left(  E/E_{P}\right)  =\left(  1+\beta\frac{E}{E_{P}}+\gamma\frac
{E^{2}}{E_{P}^{2}}\right)  \exp(-\alpha\frac{E^{2}}{E_{P}^{2}}),\qquad
g_{2}\left(  E/E_{P}\right)  =1.
\end{equation}
Once again using the general results outlined in Appendix \ref{appe}, we find
that the two integrals $I_{1}$ and $I_{2}$, given by Eqs. $\left(
\ref{I1}\right)  $-$\left(  \ref{I2}\right)  $, take the following form%

\begin{eqnarray}
I_{1,2}&=&\frac{E_{P}^{4}}{2\pi^{2}\alpha}\,{\exp\left(  -\frac{\alpha
x_{1,2}^{2}}{2}\right)  }\left[  x_{1,2}^{2}K_{1}{\left(  {\frac
{\alpha\,x_{1,2}^{2}}{2}}\right)  }+\frac{\beta\sqrt{\pi}}{4\sqrt{{\alpha}}%
}{\exp\left(  -\frac{\alpha x_{1,2}^{2}}{2}\right)  }\left(  x_{1,2}%
^{2}+{\frac{3}{2{\alpha}}}\right)  \right.
 \nonumber  \\
&&\left.  +\gamma x_{1,2}^{2}\left(  -{\frac{x_{1,2}^{2}}{2\alpha}}K_{1}\left(
{\frac{\alpha\,x_{1,2}^{2}}{2}}\right)  +\frac{x_{1,2}^{2}}{2}\left(
-K_{0}\left(  {\frac{\alpha\,x_{1,2}^{2}}{2}}\right)  -{\frac{2}{\alpha\,}%
}K_{1}\left(  {\frac{\alpha\,x_{1,2}^{2}}{2}}\right)  \right)  -{\frac
{1}{{\alpha}}}K_{1}\left(  {\frac{\alpha\,x_{1,2}^{2}}{2}}\right)  \right)
\right]  \,, \label{I12}%
\end{eqnarray}
with the definition $x_{1,2}=\sqrt{m_{1,2}^{2}\left(  r\right)  /E_{P}^{2}}$.
We now investigate the two limiting approximations, namely, the cis-planckian
regime, where $x_{i}\ll1$ ($i=1,2$), and the trans-planckian regime, where
$x_{i}\gg1$, respectively.

Regarding the cis-planckian regime, the self-sustained equation
$\left(  \ref{LoverG}\right)  $%
\begin{align}
\frac{1}{2G}\,\frac{b^{\prime}(r)}{r^{2}}  &  =\frac{E_{P}^{4}}{2{\alpha}%
\pi^{2}}\left[  2\left(  \frac{2}{{\alpha}^{2}}\,+{\frac{3\beta\sqrt{\pi}%
}{2{\alpha}^{5/2}}}-{\frac{4\gamma}{{\alpha}^{3}}}\right)  +\left(
\frac{\gamma}{{\alpha}^{2}}-{\frac{\beta\sqrt{\pi}}{2{\alpha}^{3/2}}}-\frac
{1}{{\alpha}}\right)  \left(  {{x}_{1}^{2}+{x}_{2}^{2}}\right)  \right.
\nonumber\\
&  \left.  +\,\left(  {\frac{2\ln\left(  \alpha{{x}_{1}^{2}}/4\right)
+2\ln\left(  \alpha{{x}_{2}^{2}}/4\right)  +2\gamma_{E}+1}{8}}+\frac{\gamma
}{4{\alpha}}-\,{\frac{\sqrt{\pi}\beta}{16\sqrt{\alpha}}}\right)  \left(
{{x}_{1}^{4}+{x}_{2}^{4}}\right)  \right]  +O\left(  x^{6}\right)  ,
\label{abg}%
\end{align}
where $\gamma_{E}$ is the Euler's constant. In order to simplify the analysis,
consider the following imposition
\begin{equation}
\left\{
\begin{array}
[c]{c}%
\frac{2}{{\alpha}^{2}}\,+\frac{3\beta\sqrt{\pi}}{2{\alpha}^{5/2}}%
-\frac{4\delta}{{\alpha}^{3}}{=0}\\
\\
\frac{\delta}{{\alpha}^{2}}-\frac{\beta\sqrt{\pi}}{2{\alpha}^{3/2}}-\frac
{1}{{\alpha}}=0
\end{array}
\right.  \,,
\end{equation}
which leads to the following solution
\begin{equation}
\beta=-4\sqrt{{\frac{{\alpha}}{\pi}}};\qquad\gamma=-\alpha;\qquad
\mathrm{for}\;\;\alpha\in\mathbb{R}^{+}\,.
\end{equation}
Thus, Eq. $(\ref{abg})$ simplifies to%
\begin{equation}
\frac{1}{2G}\,\frac{b^{\prime}(r)}{r^{2}}=\frac{E_{P}^{4}}{2{\alpha}\pi^{2}%
}\left[  \left(  {\frac{\ln\left(  \alpha x_{1}^{2}/4\right)  +\ln\left(
\alpha x_{2}^{2}/4\right)  +2\gamma_{E}+1}{4}}\right)  \left(  {{x}_{1}%
^{4}+{x}_{2}^{4}}\right)  \right]  . \label{blog}%
\end{equation}
Note that as the previous formula is obtained in the cis-planckian regime, the
logarithmic functions in this range are always negative and the whole
expression goes to zero from the negative side. Furthermore, as in the 
previous two examples, it is an easy matter to show that the solution 
is not asymptotically flat, and in principle can be matched to an 
exterior vacuum solution at a junction interface.

Let us now fix our attention on the trans-planckian regime, where $x_{i}\gg1$ ($i=1,2$).
The integrals in the Eq. $\left(  \ref{I12}\right)  $ have the following
asymptotic expansion%
\begin{equation}
I_{1,2}\simeq\frac{E_{P}^{4}}{2\pi^{2}}\,{\exp\left(  -\alpha x_{1,2}%
^{2}\right)  }\frac{\sqrt{\pi}}{{\alpha}^{3/2}}\left(  -{\gamma\,{x}_{1,2}%
^{3}}+{\frac{\beta\,{x}_{1,2}^{2}}{4}}+\,{\frac{\left(  4\alpha-9\gamma
\right)  x_{1,2}}{4{\alpha}}}+{\frac{3\beta}{8{\alpha}}}+{\frac{3\left(
8\alpha-15\gamma\right)  }{32{\alpha}^{2}x_{1,2}}}-\,{\frac{15\left(
4\alpha-7\gamma\right)  }{128{\alpha}^{3}{x}_{1,2}^{3}}}+O\left(  {x}%
_{1,2}^{-5}\right)  \right)  .
\end{equation}
Analogously to the previous cases, due to the Gaussian exponential, we find
that quantum fluctuations lead to a vanishing contribution, independently of
the choice of the parameters $\beta$ and $\gamma$. Therefore we conclude that
$b^{\prime}(r)=0$ leading to a constant shape function, $b(r)=r_{t}$.

Collecting the results of cases described in sections $\left(  \ref{p3a}%
,\ref{p3b}\right)  $ and $\left(  \ref{p3c}\right)  $, we conclude that the
model governed by the Gaussian Rainbow's function $g_{1}\left(  E/E_{P}%
\right)  $ and its polynomial variations lead only to a constant shape
function $b\left(  r_{t}\right)  =r_{t}$. In the next section, we consider a
different approach to the problem of the self sustained equation by fixing the
geometry of the wormhole. More specifically, we consider a variational 
approach which imposes a local analysis to the problem and restrict our attention 
to the behavior of the metric function $b(r)$ at the wormhole throat, $r_{t}$.

\section{Fixing the geometry}
\label{p4}

In this section, we outline a different approach in the context of
self-sustained wormholes in MDR. More specifically, we fix the shape function,
and therefore the geometry, and restricting our analysis to the throat we find
conditions on the specific parameter space in order to have wormhole
solutions. In the semi-classical context, we emphasize that solutions of
self-sustained wormholes were found in Refs. \cite{Garattini,semiclassWH}, by
using standard regularization and renormalization techniques. Thus, we apply
an analogous approach, however without taking into account any renormalization procedure, but
using only the deformed spacetime at the Planck scale.

To set the stage and for concreteness consider the specific choice for the
shape function $b(r)=r_{t}^{2}/r$ \cite{Morris}. Thus, from Eq. $(\ref{LoverG}%
)$, and restricting the analysis to the throat $r=r_{t}$, we find%
\begin{equation}
-\frac{1}{2G}\,\frac{1}{r_{t}^{2}g_{2}\left(  E\right)  }\,=\frac{2}{3\pi^{2}%
}\left(  I_{1}+I_{2}\right)  , \label{SSbr}%
\end{equation}
with $I_{1}$ and $I_{2}$ given by Eqs. $(\ref{I1})$ and $(\ref{I2})$,
respectively. Maintaining the setting $g_{2}(E/E_{P})=1$, we find that the
effective masses at the throat simplify to%
\begin{equation}
\left\{
\begin{array}
[c]{c}%
m_{1}^{2}\left(  r_{t}\right)  =\frac{3}{r_{t}^{2}}\\
\\
m_{2}^{2}\left(  r_{t}\right)  =-\frac{1}{r_{t}^{2}}%
\end{array}
\right.  .
\end{equation}
Therefore $I_{1}$ and $I_{2}$ become%
\begin{equation}
I_{1}=3\int_{3/r_{t}^{2}}^{\infty}g_{1}\left(  E/E_{P}\right)  E^{2}%
\sqrt{E^{2}-\frac{3}{r_{t}^{2}}}dE \label{I1a}%
\end{equation}
and%
\begin{equation}
I_{2}=3\int_{0}^{\infty}g_{1}\left(  E/E_{P}\right)  E^{2}\sqrt{E^{2}+\frac
{1}{r_{t}^{2}}}dE\,, \label{I1b}%
\end{equation}
respectively.

Now, in order to have only one solution with variables $\alpha$ and $r_{t}$,
we demand that%
\begin{equation}
\frac{d}{dr_{t}}\left[  -\frac{1}{2G}\,\frac{1}{r_{t}^{2}}\right]  \,=\frac
{d}{dr_{t}}\left[  \frac{2}{3\pi^{2}}\left(  I_{1}+I_{2}\right)  \right]  \,,
\end{equation}
which takes the following form
\begin{equation}
1=\frac{2}{\pi^{2}}\left[  3\int_{\sqrt{3}/r_{t}E_{P}}^{\infty}g_{1}\left(
u\right)  u^{2}\sqrt{u^{2}-\frac{3}{\left(  r_{t}E_{P}\right)  ^{2}}}%
du-\int_{0}^{\infty}g_{1}\left(  u\right)  u^{2}\sqrt{u^{2}+\frac{1}{\left(
r_{t}E_{P}\right)  ^{2}}}du\right]  ,
\end{equation}
where we have set $u=E/E_{P}$ and $G^{-1}=E_{P}^{2}$.

Consider a specific form of $g_{1}(u)$ given by $g_{1}(u)=\exp(-\alpha u^{2})$
with $\alpha$ variable, and after integration, we find that%
\begin{equation}
1=\frac{2}{\pi^{2}}\frac{d}{d\alpha}\left[  \frac{1}{2}\exp\left(
\frac{\alpha}{2\left(  r_{t}E_{P}\right)  ^{2}}\right)  K_{0}\left(
\frac{\alpha}{2\left(  r_{t}E_{P}\right)  ^{2}}\right)  \right]  -\frac{6}%
{\pi^{2}}\frac{d}{d\alpha}\left[  \frac{1}{2}\exp\left(  -\frac{3\alpha
}{2\left(  r_{t}E_{P}\right)  ^{2}}\right)  K_{0}\left(  \frac{3\alpha
}{2\left(  r_{t}E_{P}\right)  ^{2}}\right)  \right]  , \label{SSEq}%
\end{equation}
where $K_{0}(x)$ is a modified Bessel function of order 0. Writing the
explicit form of the derivatives in Eq. $\left(  \ref{SSEq}\right)  $ we find
\begin{equation}
1=\frac{1}{2\pi^{2}x^{2}}f\left(  \alpha,x\right)  \,, \label{SSEq1}%
\end{equation}
where, for notational simplicity, we have used $x=r_{t}E_{P}$ and used the
definition
\begin{equation}
f\left(  \alpha,x\right)  =\exp\left(  \frac{\alpha}{2x^{2}}\right)
K_{0}\left(  \frac{\alpha}{2x^{2}}\right)  -\exp\left(  \frac{\alpha}{2x^{2}%
}\right)  K_{1}\left(  \frac{\alpha}{2x^{2}}\right)  +9\exp\left(
-\frac{3\alpha}{2x^{2}}\right)  K_{0}\left(  \frac{3\alpha}{2x^{2}}\right)
+\exp\left(  -\frac{3\alpha}{2x^{2}}\right)  K_{1}\left(  \frac{3\alpha
}{2x^{2}}\right)  .
\end{equation}

The procedure now is in principle straightforward. In order to have one and
only one solution, we demand that the expression in the r.h.s. of Eq.
$(\ref{SSEq1})$ has a stationary point with respect to $x$ which coincides
with the constant value $1$. For a generic but small $\alpha$, we can expand
in powers of $\alpha$ to find%
\begin{equation}
0=\frac{d}{dx}\left[  \frac{1}{2\pi^{2}x^{2}}f\left(  \alpha,x\right)
\right]  \simeq\frac{20-10\ln\left(  4x^{2}/\alpha\right)  +10\gamma_{E}%
+9\ln3}{\pi^{2}x^{2}}+O\left(  \alpha\right)  \,,
\end{equation}
which has a root at%
\begin{equation}
\bar{x}=r_{t}E_{P}=2.973786871\sqrt{\alpha}\,.\label{xmin}%
\end{equation}
Substituting $\bar{x}$ into Eq. $(\ref{SSEq1})$, we find%
\begin{equation}
1=\frac{0.2423530631}{\alpha}\,,
\end{equation}
fixing therefore $\alpha\simeq0.242$. It is interesting to note that
this value is very close to the value $\alpha=1/4$ used in
Ref. \cite{RGGM} inspired by a non-commutative analysis \cite{RGPN}. As in Ref.
\cite{semiclassWH}, it is rather important to emphasize a shortcoming in the
analysis carried in this section, mainly due to the technical difficulties
encountered. Note that we have considered a variational approach which imposes
a local analysis to the problem, namely, we have restricted our attention to
the behavior of the metric function $b(r)$ at the wormhole throat, $r_{t}$.
Despite the fact that the behavior is unknown far from the throat, due to the
high curvature effects at or near $r_{t}$, the analysis carried out in this
section should extend to the immediate neighborhood of the wormhole throat.
Nevertheless it is interesting to observe that in Ref. \cite{Garattini} the greatest value of the 
wormhole throat was fixed at $r_{t}\simeq 1.16/E_{P}$ using a regularization-renormalization scheme. 
From Eq. $\left(  \ref{xmin}\right)$, one immediately extracts $r_{t}\simeq1.46/E_{P}$ which is 
slightly larger.

\section{Summary and discussion}
\label{p5}

Wormhole are hypothetical tunnels that violate the null energy condition, and
therefore all of the energy conditions, and thus it seems that a natural
environment of these exotic spacetimes lies in the quantum regime, as a large
number of quantum systems have been shown to violate the energy conditions \cite{Morris,Visser},
such as the Casimir effect. In this context, it has been shown that various
wormhole solutions in semi-classical gravity have been found in the literature
\cite{semiclass,Garattini,semiclassWH}. In the semi-classical
approach, the Einstein field equation takes the form $G_{\mu\nu}=8\pi
G\,\langle T_{\mu\nu}\rangle^{\mathrm{ren}}$, where the term $\langle
T_{\mu\nu}\rangle^{\mathrm{ren}}$ is the renormalized expectation value of the
stress-energy tensor operator of the quantized field. In addition to this, the metric is
separated into a background component, $\bar{g}_{\mu\nu}$ and a perturbation
$h_{\mu\nu}$, i.e., $g_{\mu\nu}=\bar{g}_{\mu\nu}+h_{\mu\nu}$. A key aspect is
that the Einstein tensor may also be separated into a part describing the
curvature due to the background geometry and that due to the perturbation,
namely, $G_{\mu\nu}(g_{\alpha\beta})=G_{\mu\nu}(\bar{g}_{\alpha\beta})+\Delta
G_{\mu\nu}(\bar{g}_{\alpha\beta},h_{\alpha\beta})$ where $\Delta G_{\mu\nu
}(\bar{g}_{\alpha\beta},h_{\alpha\beta})$ may be considered a perturbation
series in terms of $h_{\mu\nu}$. Using the semi-classical Einstein field
equation, in the absence of matter fields, the effective stress-energy tensor
for the quantum fluctuations is given by $8\pi G\,\langle T_{\mu\nu}%
\rangle^{\mathrm{ren}}=-\langle\Delta G_{\mu\nu}(\bar{g}_{\alpha\beta}%
)\rangle^{\mathrm{ren}}$ so that the equation governing the quantum
fluctuations behaves as a backreaction equation. Thus, the possibility that
wormhole geometries were sustained by their own quantum fluctuations, using
the above-mentioned semi-classical approach was shown in
\cite{Garattini,semiclassWH}. More specifically, the graviton one loop
contribution to a classical energy in a wormhole background was taken into
account, and a variational approach with Gaussian trial wave functionals was
used. A zeta function regularization was involved to handle with divergences
and a renormalization procedure was introduced and the finite one loop energy
was considered as a \textit{self-consistent} source for the traversable wormhole.

In this work, we have considered the possibility that wormhole geometries are
also sustained by their own quantum fluctuations, but in the context of
modified dispersion relations. We considered different models regulated by the
Rainbow's function $g_{1}\left(  E/E_{P}\right) $, given by Eq.
(\ref{Remofunctions}), to analyse the effect on the form of the shape function
$b(r)$, and found specific solutions for wormhole geometries in the
cis-planckian regime and trans-planckian regime. In the latter regime, we
found that the quantum correction are exponentially suppressed, thus leading
to a constant shape function, i.e., $b(r)=r_{t}$, where $r_{t}$ is the
wormhole throat. In the cis-planckian regime, the solutions found do not
represent asymptotically flat wormhole geometries. However, for these cases,
one may in principle match these interior wormhole solutions with an exterior
vacuum Schwarzschild spacetime. In addition to this analysis, we also fixed
the geometry by considering the behaviour of a specific shape function $b(r)$,
through a variational approach, at the wormhole throat. 
We explored the parameter range of the Rainbow's functions and found a good 
agreement with previous results used in the literature \cite{RGGM}. It is 
important to point out that the above-mentioned variational approach
presents a shortcoming mainly due to the technical difficulties encountered.
Note that the latter variational approach considered imposes a local
analysis to the problem, namely, we have restricted our attention to the
behavior of the metric function $b(r)$ at the wormhole throat, $r_{t}$ (this is similar to the 
analysis carried out in \cite{semiclassWH}).
Despite the fact that the behavior is unknown far from the throat, due to the
high curvature effects at or near $r_{t}$, the analysis carried out in this
context should extend to the immediate neighborhood of the wormhole throat.

\appendix{}

\section{Integrals}

\label{appe}In this Appendix we provide the rules to solve the integrals $I_{1}$
and $I_{2}$, given by Eqs. $\left(  \ref{I1}\right)  $-$\left(  \ref{I2}%
\right)  $, with the following choices for the Rainbow's functions
\begin{equation}
g_{1}(\frac{E}{E_{P}})=\sum_{i=0}^{n}\beta_{i}\frac{E^{i}}{E_{P}^{i}}%
\exp(-\alpha\frac{E^{2}}{E_{P}^{2}}) \,, \qquad g_{2}\left(  E/E_{P}\right)
=1;\qquad\alpha>0,\beta_{i}\in\mathbb{R}.
\end{equation}
In some particular cases such as in section \ref{p4}, the effective mass becomes
negative and the integral takes the form%
\begin{equation}
I_{+}=\int_{0}^{\infty}g_{1}(\frac{E}{E_{P}})E^{2}\sqrt{E^{2}+m^{2}}dE\,.
\label{I+}%
\end{equation}
Therefore we divide the integration in two classes denoted by the integrals
$I_{+}$ and $I_{-}$ described by%
\begin{equation}
I_{-}=\int_{\sqrt{m^{2}}}^{\infty}g_{1}(\frac{E}{E_{P}})E^{2}\sqrt{E^{2}%
-m^{2}}dE. \label{I-}%
\end{equation}
For the examples discussed in section \ref{p3}, the following
relationships for $I_{-}$ are extremely useful
\begin{equation}
\frac{d^{n}I_{-}}{d\alpha^{n}}=\left(  -\right)  ^{n}E_{P}^{2n}\int
_{\sqrt{m^{2}}}^{\infty}\exp(-\alpha\frac{E^{2}}{E_{P}^{2}})E^{2n}\sqrt
{E^{2}-m^{2}}dE\qquad n\geq1
\end{equation}
and the following for $I_{+}$%
\begin{equation}
\frac{d^{n}I_{+}}{d\alpha^{n}}=\left(  -\right)  ^{n}E_{P}^{2n}\int
_{0}^{\infty}\exp(-\alpha\frac{E^{2}}{E_{P}^{2}})E^{2n}\sqrt{E^{2}+m^{2}%
}dE\qquad n\geq1.
\end{equation}

Taking into account $g_{1}(\frac{E}{E_{P}})=\exp(-\alpha\frac{E^{2}
}{E_{P}^{2}})$ in $I_{-}$, we find that changing variables $E=\sqrt{x}$, we obtain%
\begin{align}
I_{-}  &  =\frac{1}{2}\int_{\sqrt{m^{2}}}^{\infty}\exp(-\alpha\frac{x}%
{E_{P}^{2}})\sqrt{x}\sqrt{x-m^{2}}dx\nonumber\\
&  =\frac{E_{P}^{4}}{2\sqrt{\pi}}\left(  \frac{m^{2}}{\alpha E_{P}^{2}%
}\right)  \Gamma\left(  \frac{3}{2}\right)  \exp\left(  -\frac{\alpha m^{2}%
}{2E_{P}^{2}}\right)  K_{1}\left(  \frac{\alpha m^{2}}{2E_{P}^{2}}\right)  ,
\end{align}
where we have used the following relationship
\begin{equation}
\int_{u}^{\infty}\left(  x-u\right)  ^{\mu-1}x^{\mu-1}\exp\left(  -\beta
x\right)  dx=\frac{1}{\sqrt{\pi}}\left(  \frac{u}{\beta}\right)  ^{\mu
-1/2}\Gamma\left(  \mu\right)  \exp\left(  -\frac{\beta u}{2}\right)
K_{\mu-1/2}\left(  \frac{\beta u}{2}\right)  \qquad%
\begin{array}
[c]{c}%
\mathrm{Re}\,\mu>0\\
\mathrm{Re}\,\beta u>0
\end{array}
.
\end{equation}

The same argument applies for $I_{+}$ to obtain%
\begin{align}
I_{+}  &  =\frac{1}{2}\int_{0}^{\infty}\exp(-\alpha\frac{x}{E_{P}^{2}}%
)\sqrt{x}\sqrt{x+m^{2}}dx\nonumber\\
&  =\frac{E_{P}^{4}}{2\sqrt{\pi}}\left(  \frac{m^{2}}{\alpha E_{P}^{2}%
}\right)  \Gamma\left(  \frac{3}{2}\right)  \exp\left(  \frac{\alpha m^{2}%
}{2E_{P}^{2}}\right)  K_{-1}\left(  \frac{\alpha m^{2}}{2E_{P}^{2}}\right)  ,
\end{align}
where we have used the following relationship%
\begin{equation}
\int_{0}^{\infty}\left(  x+\beta\right)  ^{\nu-1}x^{\nu-1}\exp\left(  -\mu
x\right)  dx=\frac{1}{\sqrt{\pi}}\left(  \frac{\beta}{\mu}\right)  ^{\nu
-1/2}\Gamma\left(  \nu\right)  \exp\left(  \frac{\beta\mu}{2}\right)
K_{1/2-\nu}\left(  \frac{\beta\mu}{2}\right)  \qquad%
\begin{array}
[c]{c}%
\mathrm{Re}\,\mu>0\\
\mathrm{Re}\,\nu>0\\
\left\vert \arg\beta\right\vert <\pi
\end{array}
.
\end{equation}

On the other hand, the integrals of the form%
\begin{equation}
\frac{d^{n}I_{-}^{o}}{d\alpha^{n}}=\left(  -\right)  ^{n}E_{P}^{2n}\int
_{\sqrt{m^{2}}}^{\infty}\exp(-\alpha\frac{E^{2}}{E_{P}^{2}})E^{2n+1}%
\sqrt{E^{2}-m^{2}}dE\qquad n\geq2 \,,
\end{equation}
can be generated by the following expression%
\begin{equation}
\int_{a}^{\infty}dx\left(  x-a\right)  ^{1/2}x\exp\left(  -\mu x\right)
=\frac{\sqrt{\pi}}{4}\mu^{-5/2}(3+2\mu a)\exp\left(  -\mu a\right)  \qquad
a>0,\mu>0,
\end{equation}
while integrals of the form%
\begin{equation}
\frac{d^{n}I_{+}^{o}}{d\alpha^{n}}=\left(  -\right)  ^{n}E_{P}^{2n}\int
_{0}^{\infty}\exp(-\alpha\frac{E^{2}}{E_{P}^{2}})E^{2n+1}\sqrt{E^{2}+m^{2}%
}dE\qquad n\geq2
\end{equation}
can be generated by the following relationship%
\begin{equation}
\int_{0}^{\infty}dx\left(  x+t\right)  ^{1/2}x\exp\left(  -\mu x\right)
=\frac{3}{2}\frac{\sqrt{t}}{\mu^{2}}+\frac{\sqrt{\pi}}{4}\mu^{-5/2}\exp
(t\mu)(3-2t\mu)\text{Erfc}\left[  \sqrt{t\mu}\right]  \qquad t>0,\mu>0.
\end{equation}

\acknowledgments

FSNL acknowledges financial support of the Funda\c{c}\~{a}o para a Ci\^{e}ncia
e Tecnologia through Grants PTDC/FIS/102742/2008 and CERN/FP/116398/2010.



\end{document}